# Mise en place d'un protocole d'identification de traits articulatoires communs à la gestualité co-verbale du français et à la Langue des Signes Française (LSF) : application au geste épistémique

Fanny Catteau°
Claudia S. Bianchini*

RESUME: Cet article se focalise sur les caractéristiques articulatoires des gestes épistémiques (i.e. des gestes produits lors de l'expression de la certitude ou de l'incertitude) en gestualité co—Verbale (GCV) du français et en Langue des Signes Français (LSF). Il présente une nouvelle méthodologie d'analyse qui repose sur l'utilisation complémentaire d'une annotation manuelle (réalisée avec le système de transcription Typannot) et d'une annotation semi—automatique (via un logiciel d'estimation de poses, AlphaPose) afin de mettre en évidence les caractéristiques kinésiologiques de ces gestes épistémiques. La méthodologie présentée permet notamment d'analyser les positions et les mouvements de flexion/extension du cou en contexte d'épistémicité; les résultats de cette analyse montrent qu'en GCV du français et en LSF: (1) des hochements de tête passant par la position neutre (c'est-à-dire, tête droite, sans flexion/extension) et une vitesse de mouvement élevée sont des marqueurs de certitude; et (2) les tenues de position de la tête hors de la position neutre et une faible vitesse de mouvement indiquent l'incertitude. Cette étude est menée dans le cadre du projet ANR LexiKHuM qui développe des solutions de communication kinesthésique pour l'interaction humain machine.

ABSTRACT: This article focuses on the articulatory characteristics of epistemic gestures (i.e., gestures used to express certainty or uncertainty) in co—speech gestures (CSG) in French and in French Sign Language (LSF). It presents a new methodology for analysis, which relies on the complementary use of manual annotation (using Typannot) and semi—automatic annotation (using AlphaPose) to highlight the kinesiological characteristics of these epistemic gestures. The presented methodology allows to analyze the flexion/extension movements of the head in epistemic contexts. The results of this analysis show that in CSG and LSF: (1) head nods passing through the neutral position (i.e., head straight with no flexion/extension) and high movement speed are markers of certainty; and (2) holding the head position away from the neutral position and low movement speed indicate uncertainty. This study is conducted within the framework of the ANR LexiKHuM project, which develops kinesthetic communication solutions for human—machine interaction

---

° Université de Poitiers - Laboratoire UR15076-FORELLIS. courriel : funnycatteau@orange.fr
* Université de Poitiers - Laboratoire UR15076-FORELLIS.
   courriel : claudia.savina.bianchini@univ-poitiers.fr – auteur correspondant

**Introduction**

Dans les corpus multimodaux de langues vocales ou de langues des signes (LS), les gestes et les signes sont plus souvent identifiés par leur fonction que par leur forme (Chevrefils *et al.*, 2021). Même lorsque la forme est prise en compte, le mouvement n'est pas vu comme le résultat d'un changement au niveau des articulations des segments corporels, mais comme le résultat d'un déplacement le long d'une trajectoire (Boutet, 2018). Ainsi, il sera plus probable de trouver une annotation indiquant que la main a tracé une grande ligne allant de droite à gauche plutôt que de lire une description de ce que le bras et l'avant-bras ont dû faire pour permettre à la main de tracer cette ligne.

Il est donc rare, dans les corpus multimodaux, de trouver une analyse articulatoire des mouvements corporels, et cela est dû à : la difficulté de définir un modèle articulatoire suffisamment détaillé pour pouvoir réaliser une analyse fine des dynamiques du mouvement ; la difficulté de produire des données qui puissent être manipulées et donc analysées ; la difficulté de définir un processus d'annotation des données qui allie finesse descriptive et rapidité, afin de pouvoir coder assez de données pour mener des analyses quantitatives.

Cet article vise à fournir une réponse théorique et méthodologique à ces trois problèmes, en proposant l'adoption d'un modèle kinésiologique de description du mouvement (Boutet, 2018) et en lui associant Typannot (Bianchini *et al.*, 2018 ; Bianchini, 2024), un système de transcription facilitant la manipulation des données, et AlphaPose (Fang *et al.*, 2023), un logiciel de détection semi-automatisée des postures corporelles et du mouvement. Le protocole ainsi défini a été appliqué à une étude dont l'objectif est d'analyser les traits articulatoires des expressions épistémiques (certitude et incertitude) communs à la gestualité co-verbale (GcV) du français et à la langue des signes française (LSF). Cette recherche s'inscrit dans le cadre du projet LexiKHuM (Lexique Kinésiologique d'interaction Humain-Machine[1]), financé par l'Agence Nationale de la Recherche (ANR), qui vise à améliorer la communication humain-machine à travers la prise en compte de la dimension épistémique de la communication.

La première partie de l'article pose le cadre théorique en présentant la notion d'épistémicité, les raisons pour lesquelles il est important de la prendre en compte dans la communication humain-machine et la manière dont elle a lieu dans la communication humain-humain. Cette première partie contient les hypothèses qui seront testées grâce à la mise en place de notre protocole de description et de codage semi-automatique qui est présenté dans la deuxième partie de l'article. Enfin, la troisième partie présente les résultats préliminaires de l'application du protocole à un corpus de GCV du français et de LSF. Suivent les conclusions et les perspectives de ce travail.

---

[1] ANR–LexiKHuM (2021-2025). https://anr.fr/Projet-ANR-20-CE33-0012 (08/10/23).

1. **Geste épistémique et interaction**

*1.1. Interaction humain-machine et épistémicité*

Dans un monde de plus en plus automatisé, un grand nombre de tâches sont désormais réalisées conjointement par un humain et une machine dotée d'un *système intelligent* (SI) : un dispositif électronique et informatique qui peut exécuter des tâches avec l'humain ou à sa place, mais qui peut aussi conseiller à l'humain, voire lui imposer, la manière d'exécuter une action. Ces SI sont en mesure de recueillir une information, de la traiter, et ensuite de produire une instruction, une action ou même une nouvelle information qui sera traitée par l'humain ou par le SI lui-même. Bien que leur utilité soit réelle, les journaux rapportent souvent la nouvelle d'accidents survenus lors de l'utilisation de ces SI : il peut s'agir de problèmes liés à la manière dont ils ont été programmés, mais il est fréquent que les accidents soient causés par des problèmes de communication entre l'opérateur humain et le SI. La compréhension et la compensation de ces difficultés d'interaction entre humain et machine constituent donc un enjeu de recherche majeur pour les ingénieurs et les roboticiens qui développent ces SI.

Ces chercheurs ont identifié dans l'*opacité* des SI une des causes majeures des difficultés de communication entre SI et humains (Christoffersen & Woods, 2002 ; Dekker & Woods, 2002) : les SI sont faits pour traiter des informations mais la manière dont ils les traitent est *opaque*, peu accessible aux humains qui participent à l'interaction. Ainsi, au cours de son fonctionnement, un SI peut être amené à déclencher *une réaction en cascade* : il traite une information à laquelle il réagit en produisant une autre information, qui provoque en réaction la production d'une nouvelle information, et ainsi de suite jusqu'à ce que la réaction en chaîne s'arrête. Il suffit d'une information erronée ou mal interprétée par le SI, pour que le résultat final devienne imprévisible (Taleb, 2012), avec des conséquences parfois importantes. Le manque de visibilité sur la manière dont les informations sont traitées au cours du processus diminue, voire élimine, la capacité de l'opérateur humain de comprendre la réaction en cours. Cela l'empêche de prédire le fonctionnement du SI (Grynszpan *et al.*, 2019), et par conséquent limite sa capacité d'intervenir au milieu du processus pour corriger les informations.

L'une des raisons de la difficulté à transmettre leurs intentions, semble être que les SI ne prennent généralement pas en compte la *dimension épistémique de la communication* : c'est-à-dire la posture qu'a le participant à une interaction par rapport au *degré de certitude ou d'incertitude* des connaissances qu'il possède et des connaissances qu'il attribue à son interlocuteur. Les SI n'informent pas l'opérateur humain de leur incertitude face à la fiabilité d'une information reçue ou produite (Pezzulo *et al.*, 2021).

La nécessité de connaître la fiabilité des informations reçues et d'informer sur la fiabilité des informations fournies n'est pas propre à l'interaction humain-machine : lorsqu'ils communiquent entre eux, les êtres humains fournissent constamment à leur interlocuteur des informations concernant leur engagement et leurs sentiments de certitude et d'incertitude à l'égard des propositions formulées (Roseano *et al.*, 2016). L'étude approfondie de la dimension épistémique dans l'interaction humain-humain ouvre la voie à son intégration dans l'interaction humain-machine, réduisant ainsi l'opacité des SI tout en rendant plus prévisible leur comportement (Pezzulo *et al.*, 2021), ce qui permettrait à l'opérateur humain d'anticiper les résultats des réactions en cascade et d'éviter les accidents.

Un projet interdisciplinaire – reposant sur la collaboration de linguistes, roboticiens et experts de sciences cognitives – a donc vu le jour pour répondre à cet enjeu. Le projet LexiKHuM propose de développer un SI qui prenne en compte la dimension épistémique de la communication, en lui fournissant le lexique pour exprimer son degré de certitude ou d'incertitude à propos des informations qu'il reçoit de l'extérieur et/ou qu'il restitue à son opérateur humain.

LexiKHuM se base sur trois hypothèses : il est possible d'améliorer la communication des intentions des SI à travers la prise en compte de la dimension épistémique ; il est possible de reprendre des caractéristiques de la communication humaine de l'épistémicité pour les implémenter dans un SI ; plus grande sera la ressemblance entre le langage naturel et le lexique conçu pour le projet, plus aisée sera la compréhension du message du SI de la part de l'opérateur humain.

### 1.2. *Interaction humain-humain et épistémicité*

Dans la communication humaine, la dimension sémantico-pragmatique des énoncés est le plus souvent l'indice majeur permettant d'identifier un énoncé épistémique. Néanmoins, d'autres marqueurs ont été observés par les linguistes : en effet des langues vocales (LV) comme le français, l'anglais, l'allemand ou l'espagnol présentent des spécificités à l'interface morphosyntaxique et sémantique (à travers, par exemple, le recours à des verbes modaux ou des adverbes de certitude/incertitude ; l'utilisation de termes comme « sûr », « certain », « incertain » en français) en lien avec l'expression de l'épistémicité (de Haan, 2005 ; Bross, 2012 ; Kronning, 2012). Ces mêmes marqueurs sont repérés aussi dans des langues n'ayant pas recours au canal audio-phonatoire, c'est-à-dire les LS : la présence de signes lexicaux indicateurs d'incertitude comme le signe signifiant « peut-être » ou « il semblerait » sont relevés par Herrmann (2013) en LS allemande (DGS) et par Gianfreda *et al.* (2014) en LS italienne (LIS).

Ces marqueurs lexicaux pourraient constituer une base pour la création d'un lexique d'interaction épistémique entre humain et SI : intégrés à la programmation du SI, ils pourraient être reproduits par celui-ci afin que l'humain puisse percevoir la certitude ou l'incertitude du système, tout comme il le ferait dans une interaction humain-humain. Toutefois, le secteur d'application souhaité pour le SI développé au sein de LexiKHuM est le pilotage d'hélicoptères assisté par ordinateur : le lexique développé doit utiliser un canal de communication n'étant pas déjà surchargé, ce qui exclut de pouvoir recourir à l'ouïe ou à la vue du pilote pour lui transmettre des informations concernant la certitude ou l'incertitude du SI qui l'assiste dans le pilotage. Ce besoin de « décharger » les voies de communication classiques est présent aussi dans de multiples autres scénarios d'interaction humain-machine puisque les opérateurs humains réalisent souvent des tâches complexes lorsqu'ils ont recours à ces SI. La volonté de ne pas recourir à la vue et à l'ouïe pousse à rechercher d'autres voies de communication disponibles : le goût et l'odorat n'étant pas très utilisés dans la communication langagière, le choix a été fait d'explorer la voie kinesthésique, qui englobe tant le toucher que la perception des postures et mouvements de parties de son propre corps et, au sens large, du corps de l'autre — grâce aux « mécanismes de résonance motrice entre les mouvements d'un agent et le répertoire moteur d'un observateur » (Guillain & Pry, 2012). LexiKHuM se fonde donc aussi sur deux autres hypothèses : il est possible de ne pas recourir à l'ouïe et à la vue de l'opérateur pour lui transmettre les informations concernant le degré de certitude et d'incertitude du SI ; il est possible d'implémenter un lexique basé sur la voie kinesthésique dans un SI si celui-ci est doté d'une interface haptique[2] en mesure de reproduire les caractéristiques de la communication kinesthésique.

Dans l'interaction humain-humain, la voie kinesthésique au sens large est particulièrement sollicitée dans la communication gestuelle, qu'il s'agisse de GcV des entendants, de LS utilisées par les sourds, ou de communication haptique ou de LS tactile utilisée par les personnes sourdes-aveugles. Bien que les études concernant la gestualité épistémique soient plus rares, il est possible d'analyser les caractéristiques communes à ces différentes formes de communication afin de déterminer les caractéristiques qui pourraient constituer la base d'un lexique de communication kinesthésique humain-machine. Ainsi, Roseano *et al.* (2016), dans leur étude sur l'expression prosodique et multimodale de l'épistémicité, expliquent que – dans diverses LV comme le français et l'anglais – les hochements de tête marquent la certitude, et le haussement des épaules et la moue de la bouche sont des marqueurs d'incertitude. Borràs-Comes *et al.* (2011 ; 2019) indiquent également qu'en catalan, l'incertitude est caractérisée par l'inclinaison de la tête, la secousse de la tête et/ou le froncement des sourcils. Enfin, Ferré (2012), Debras (2017) et Boutet *et al.* (2021), décrivent le *shrug*[3] (Figure 1) en français et en anglais utilisé en contexte d'incertitude. Pour les LS, les marqueurs du geste épistémique ont été décrits, entre autres, par Herrmann (2013) et Karabüklü *et al.* (2018). Ces études montrent que les signeurs exprimant de l'incertitude en DGS et en LS turque (TİD) présentent de lents hochements et une inclinaison de la tête, le froncement du front et des sourcils, les yeux plissés et/ou le regard vague, les commissures des lèvres orientées vers le bas et le buste incliné vers l'arrière. En contexte de certitude, en revanche, les signeurs produisent des hochements de tête rapides et ont les yeux ouverts. Il est intéressant de noter que la variation de la vitesse et de l'amplitude des mouvements font partie des paramètres prosodiques des gestes des LS (Wilbur, 1999) et des GcV[4].

---

[2] Une interface haptique est un dispositif technologique permettant à l'utilisateur de ressentir des sensations tactiles (grâce à des vibrations, des pressions et/ou des mouvements) lorsqu'il interagit avec un système informatique.

[3] Le *shrug* est un geste qui, dans sa réalisation la plus complexe, est caractérisé par des mains plates, dont les paumes sont orientées vers le haut (position communément appelée *palm-up*), un haussement des sourcils et des épaules, l'inclinaison de la tête et une moue de la bouche.

[4] Voir Martel *et al.* (2022 ; dans ce volume) pour une caractérisation de la prosodie des gestes de la LSF et du français parlé.

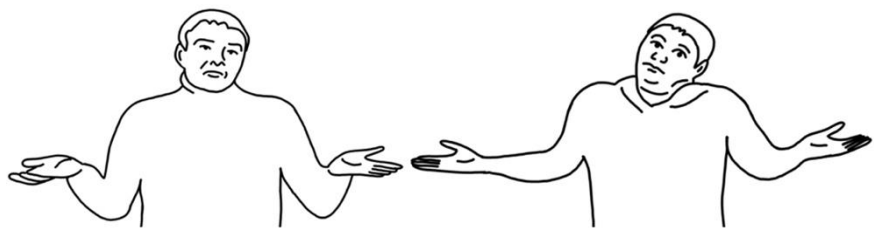

*Figure 1 –Exemples de shrugs dits 'complexes', mobilisant plusieurs articulateurs : les membres supérieurs, les épaules et la tête (image réélaborée à partir de Debras, 2017)*

L'analyse de la littérature met en évidence les caractéristiques, très souvent communes GCV et en LS, des gestes épistémiques (Tableau 1). Toutefois, dans cette littérature sur le geste épistémique, les marqueurs de l'épistémicité sont présentés de manière holistique[5] : les auteurs mentionnent les articulateurs pertinents à observer et leurs positions, mais cette description n'est souvent pas accompagnée d'une analyse fine des caractéristiques articulatoire des mouvements de ces articulateurs.

| articulateurs | CERTITUDE | INCERTITUDE |
|---|---|---|
| membres supérieurs (notamment les mains) | X | palm-up longue durée |
| tête et cou | hochements clairs et rapides | inclinaison sur le côté, hochements lents |
| épaules | X | haussement |
| buste | X | Inclinaison en arrière |
| sourcils | haussements (yeux ouverts) | froncements |
| bouche | X | moue |
| regard | X | vague |

*Tableau 1 – Synthèse des marqueurs épistémiques recensés en littérature pour la gestualité co-verbale et les langues des signes*

Le besoin d'identifier et de synthétiser les caractéristiques des gestes pour pouvoir les reproduire dans un dispositif haptique nécessite une description plus détaillée des dynamiques du mouvement que celles proposées en littérature. Il a donc été décidé de suivre ici une approche descriptive de type kinésiologique (Boutet, 2018 ; Chevrefils *et al.*, 2021), basée sur la prise en compte des variations de *degrés de liberté* (DDL) de chaque segment corporel (mains, avant-bras, bras, épaule, buste, cou, tête). Boutet (2018) définit les DDL comme « une direction de mouvement selon deux sens opposés (par exemple la flexion et l'extension) ; ils sont tous caractérisés par un axe de rotation repéré au niveau d'une articulation […], parfois sur deux articulations. Autour de chacun de ces axes, un segment tourne et est systématiquement encadré par une amplitude » (Boutet, 2008). Les DdL qu'il retient, et qui sont utilisée ici, sont la flexion/extension (FlxExt), l'abduction/adduction (AbdAdd), la rotation interne/externe (RinRex) de chaque segment participant au déploiement du geste.

---

[5] Sauf pour Boutet *et al.* (2021).

Par cette approche nous avons testé les hypothèses suivantes : il est possible d'identifier dans le mouvement des traits articulatoires porteurs du sens épistémique qui permettent de distinguer la certitude de l'incertitude en LS et en GcV ; certains de ces traits articulatoires identifiés sont communs à la GcV et à la LS.

Les analyses présentées dans cet article concernent les mouvements du cou car la revue de littérature effectuée montre qu'il existe un net contraste entre les mouvements du cou utilisés pour marquer la certitude et l'incertitude. De plus, le cou est rarement au centre de l'intérêt des linguistes travaillant sur l'épistémicité : il est donc intéressant d'approfondir son rôle. Enfin, le dispositif haptique de LexiKHuM est un manche à retour d'effort qui peut se déplacer selon les axes avant-arrière et droite-gauche et transmettre des sensations vibratoires : du point de vue mécanique, le manche et le cou partagent de nombreuses caractéristiques communes, ce qui augmente les probabilités d'arriver à transférer les caractéristiques articulatoires du mouvement du cou sur le dispositif haptique[6].

## 2. Méthode d'analyse kinésiologique de l'épistémicité

### 2.1. *Corpus analysé*

Pour analyser les gestes épistémiques en GcV et en LS, nous avions besoin d'un corpus vidéo comparable de LV et de LS : la méthodologie présentée ci-dessous s'appliquant à tout corpus contenant ce type de données. Nous avons donc choisi d'exploiter un corpus déjà existant[7], le corpus comparable DEGELS1[8] (Braffort & Boutora, 2012), qui est constitué de données vidéos en LS française (LSF) et en français oral. Les données sont issues d'enregistrements de *maptasks* de trois dyades parlant en français (un modérateur et trois interlocuteurs, soit quatre locuteurs du français), et de trois dyades signant en LSF (deux modérateurs sourds et trois interlocuteurs sourds, soit cinq locuteurs de la LSF). Les discussions des six dyades durent entre 10 et 15 min et portent sur la description de la ville de Marseille.

---

[6] Ce transfert devant être assuré, dans un deuxième temps, par l'équipe d'ingénieurs roboticiens travaillant dans le projet.

[7] Le choix de réutiliser un corpus existant s'inscrit dans la démarche de la Sciences Ouverte, qui invite tant à rendre les données réutilisables qu'à les réutiliser quand cela est possible.

[8] https://www.ortolang.fr/market/corpora/degels1 (08/10/23)

## 2.2. *Protocole d'analyse*

Afin de pouvoir comparer les gestes épistémiques en GCV du français et en LSF, et d'identifier les traits articulatoires partagés par ces deux types de gestes, ont été créé un protocole et une grille d'analyse communs aux six dyades observées, indépendamment de leurs langues. Ce protocole et cette grille pourraient être utilisés pour étudier d'autres LV et LS.

La première étape du travail a consisté à faire une analyse sémantico-pragmatique des vidéos (s'appuyant sur la structure informationnelle des énoncés, le sens des propos formulés et les intentions perçues des locuteurs), ainsi qu'à identifier et à étiqueter – grâce au logiciel d'annotation manuelle ELAN[9] (Crasborn & Sloetjes, 2008) – les passages de certitude et d'incertitude dans le corpus. La Figure 2 et la Figure 3 présentent deux exemples de séquences de certitude sélectionnées pour cette analyse.

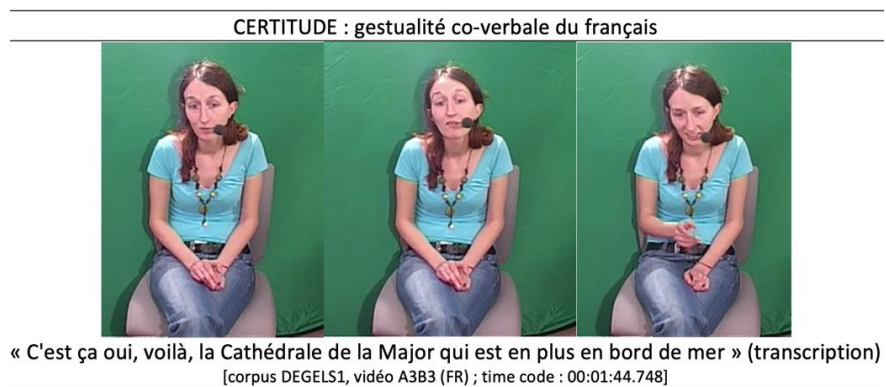

*Figure 2 – Exemple de certitude en gestualité co-verbale du français*

---

[9] https://archive.mpi.nl/tla/elan (08/10/23)

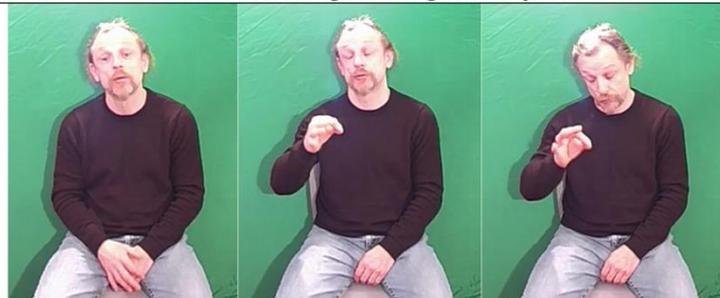

*Figure 3 – Exemple de certitude en Langue des Signes Française*

L'identification des extraits porteurs d'épistémicité ont été soumis à un accord inter-annotateur, garantissant la fiabilité de l'échantillon analysé : 40 séquences gestuelles exprimant la certitude ou l'incertitude ont été ainsi retenues, 20 séquences de certitude et 20 d'incertitude, réparties de manière homogène entre la LSF et le français.

Les gestes épistémiques sélectionnés ont ensuite été analysés en appliquant deux méthodes complémentaires : (1) une analyse articulatoire des mouvements et des positions du cou des locuteurs, reposant sur la description de ces gestes à l'œil nu à l'aide du logiciel ELAN (voir §2.3) et (2) une analyse semi-automatisée des positions et des mouvements du cou grâce aux mesures générées par AlphaPose (Fang *et al.*, 2023), un logiciel d'estimation de pose (voir §2.4).

### *2.3. Outils de représentation et description du mouvement*

La grille d'annotation utilisée pour l'analyse manuelle est conçue pour identifier et baliser les mouvements des segments corporels selon leurs variations de positions et la spécification de certaines qualités du mouvement (répétition, variation de vitesse et d'amplitude). Cette grille est inspirée de l'approche kinésiologique de Boutet (2018), qui met au cœur de l'analyse la description articulatoire des mouvements des différents segments corporels impliqués dans la réalisation des gestes (tête, cou, buste, épaules, etc.), en les décrivant en termes de variations des DDL de chaque segment (Chevrefils *et al.*, 2021).

Pour chaque séquence retenue pour l'analyse, les positions des segments composant le geste épistémique ont donc été décrites selon les variations de leurs différents DDL. Cette annotation a été réalisée avec la police de caractères Typannot (Bianchini *et al.*, 2018 ; Bianchini, 2024 ; Tableau 2), un système de transcription typographique dont le développement s'inscrit dans l'approche kinésiologique de Boutet.

*Tableau 2 – Transcription avec Typannot des positions des segments pertinents pour le projet LexiKHuM. Légende : F<small>LX</small>E<small>XT</small> = Flexion-Extension ; A<small>BD</small>A<small>DD</small> = Abduction-Adduction ; R<small>IN</small>R<small>EX</small> = Rotation interne-externe ; A<small>BD</small> D = Abduction à droite ; A<small>BD</small> G = Abduction à gauche ; R<small>IN</small> D = Rotation à droite ; R<small>IN</small> G = Rotation à gauche*

Typannot représente de manière détaillée les caractéristiques articulatoires des gestes analysés, en utilisant un vocabulaire basé sur la description des D<small>D</small>L et commun à tous les segments pris en compte. Le système subdivise l'éventail de variations des différents D<small>D</small>L en crans, ce qui discrétise les données du mouvement et facilite l'analyse du flux gestuel Typannot est composé de *caractères génériques* (Bianchini, 2024), caractères typographiques visuellement constitués d'un dessin situé dans un carré (par exemple : ⌶ qui représente le cou ou ⌂ qui représente la F<small>LX</small>E<small>XT</small>), qui représentent chacun une information nécessaire à la représentation des variations de D<small>D</small>L des différents segments corporels. Ces caractères, indépendants de toute langue vocale, facilitent la régularité et la cohérence des annotations, tout en simplifiant le processus de transcription, de lecture et de requête des annotations réalisées.

Typannot permet donc de transcrire :

- les segments corporels impliqués dans le mouvement : par exemple le cou ⍿ (l'articulateur au centre de cet article), la tête[10] ⊕, les épaules ⊤ et le buste ⌇ ; dans notre article, seules les analyses relatives au cou sont présentées ;
- les DdL attribués à chacun de ces segments : la FlxExt ⊓, l'AbdAdd ⊥ et RinRex ⊚ ; pour l'AbdAdd et la RinRex du cou et du buste, est aussi précisé le côté gauche ⇐ ou droit ⇒ ;
- les positions des segments sur chaque DdL[11] : le système permet de coder la position de repos ⊡ et les butées articulaires de chaque DdL (Flx ⊓⊙ et Ext ⊓●; Abd ⊥⊙ et Add ⊥● ; Abd à gauche ⊥⇐⊙ et Abd à droite ⊥⇒⊙ ; Rin ⊚⊙ et Rex ⊚● ; Rin à gauche ⊚⇐⊙ et Rin à droite ⊚⇒⊙), mais aussi des crans intermédiaires (grand ◐ ◑, moyen ◓ ◒, ou petit ◔ ◕[12]).

L'étude des seules variations de DdL des segments n'étant pas suffisante pour l'identification et l'analyse fine des traits articulatoires, il a été nécessaire de spécifier d'autres caractéristiques permettant de qualifier les mouvements. Pour cela, nous nous sommes inspirées des méthodologies d'étude du mouvement utilisées en analyse de la prosodie des LS (Boyes-Braem, 1999 ; Puupponen *et al.*, 2015 ; Wilbur & Malaia, 2018). En effet, la prosodie des LS est portée par le mouvement des différents segments mobilisés dans la production langagière (du buste, des mains, de la tête, des sourcils, par exemple). La durée, la vitesse et l'amplitude[13] de ces mouvements forment des contrastes dans le flux gestuel, qui créent des phénomènes prosodiques, contribuant par exemple à la focalisation (voir Martel *et al.*, 2022 — dans ce volume). Ainsi pour vérifier la présence de marqueurs épistémiques portés par les mouvements du cou, les répétitions et les contrastes d'amplitude et de vitesse de chacun de ses mouvements ont été annotés manuellement.

---

[10] Les mouvements de la tête et du cou sont souvent confondus car les variations de DdL du cou engendrent des déplacements de la tête ; ici, il a été décidé de faire une distinction nette entre les mouvements de ces deux segments. Les rotations et les inclinaisons de la tête sont toujours imputées à des variations de DdL au niveau du cou ; les hochements haut-bas de la tête peuvent dériver d'une variation de DdL du cou ou d'une variation de DdL de la tête (dans ce dernier cas, seulement si le mouvement est également possible avec un collier cervical).

[11] La position est exprimée en subdivisant en crans la mesure d'angle maximale d'un DdL, c'est-à-dire l'angle parcouru par un segment lorsqu'il passe d'une butée (par exemple la Flx maximale) à l'autre (par exemple l'Ext maximale) de ce DdL.

[12] La coloration des crans intermédiaires (noirs ◑ ◒ ◕ ou blancs ◐ ◓ ◔) dépend de la coloration (noir ● ou blanc ⊙) attribuée à la butée articulaire.

[13] Les trois paramètres prosodiques des LS (Wilbur, 1999).

La police de caractère Typannot étant toujours en cours de développement, celle-ci ne permet pas encore d'indiquer la structure temporelle et rythmique associée à ces variations de DdL. C'est pourquoi une description *ad hoc* a été développée pour les exigences de LexiKHuM : « v » représente la vitesse, « a » l'amplitude, et les symboles « + » ou « - » les contrastes, et « x1 », « x2 », « xn » donnent des indications sur les répétitions.

### 2.4. *Outils d'analyse semi-automatique des données, via AlphaPose*

Malgré le protocole d'annotation présenté ci-dessus et la mise en place de vérifications croisées entre les annotateurs, certains éléments restent complexes à coder à l'œil nu : ainsi, par exemple, il est difficile de distinguer une « petite » d'une « moyenne » FLX du cou ou d'appréhender avec précision la vitesse des mouvements. De plus, la procédure de transcription est très chronophage : la transcription par F. Catteau des 40 extraits, pour un total de 2'58", a demandé environ une semaine de travail.

Ces problématiques peuvent être résolues grâce à la mise en place d'un protocole d'analyse assistée par ordinateur, en ayant recours à des dispositifs de capture de mouvement (MoCap) ou d'estimation de pose (PosEst en mesure de produire des transcriptions semi-automatiques des vidéos. Ce type de protocole a été, entre autres, utilisé pour réaliser des analyses de la LS (cf. Jantunen *et al.*, 2016 ; Catteau, 2020 ; Chevrefils, 2022 ; Thomas, en cours).

La grande variété de dispositifs de MoCap/PosEst produisant des données numériques associées au déplacement des segments corporels permet de choisir celui qui est le plus adéquat aux exigences (et aux moyens financiers et techniques) du chercheur[14]. Dans la plupart des cas, l'extraction automatique des mesures de mouvement impose d'utiliser des dispositifs de MoCap dès l'enregistrement des données : les locuteurs doivent être équipés de capteurs et/ou être filmés avec des caméras non ordinaires (par exemple dotées d'optiques sensibles aux infrarouges) pour que les données soient exploitables. Dans le cas de corpus où l'utilisation d'outils de MoCap n'a pas été prévue en amont (tels que le corpus DEGELS1), il est possible d'utiliser des logiciels de PosEst, qui permettent d'extraire des mesures à partir de toute vidéo ayant une qualité de l'image suffisante. Le choix d'utiliser un dispositif de MoCap/PosEst plutôt qu'un autre peut influencer les caractéristiques des données extraites par ces systèmes (qui pourront être plus ou moins fiables, précises, etc.), mais non la méthodologie générale d'analyse de ces données qui est proposée ci-dessus.

---

[14] Voir Catteau (2020) et Chevrefils (2022) pour un état de l'art des dispositifs.

Pour l'étude de l'épistémicité proposée ici, nous avons utilisé AlphaPose (Fang *et al.*, 2023), un PosEst qui calcule les positions des sujets enregistrés image par image, reconstituant leur squelette à partir de la vidéo 2D. La collaboration avec des ingénieurs informatiques[15] a permis de rendre interprétables les données issues de AlphaPose, en les visualisant sous la forme de courbes tracées dans un plan cartésien dont l'abscisse représente le temps et l'ordonnée les degrés d'amplitude des DdL. Le Graphique 1a et le Graphique 2a montrent un exemple de mesure de la FlxExt du cou sur une séquence de certitude en GCV et en LSF (ces séquences sont les mêmes que celles représentées dans la Figure 2 et la Figure 3) ; le Graphique 1b et le Graphique 2b représentent la vitesse des mouvements sur l'axe de FlxExt du cou de ces deux mêmes extraits.

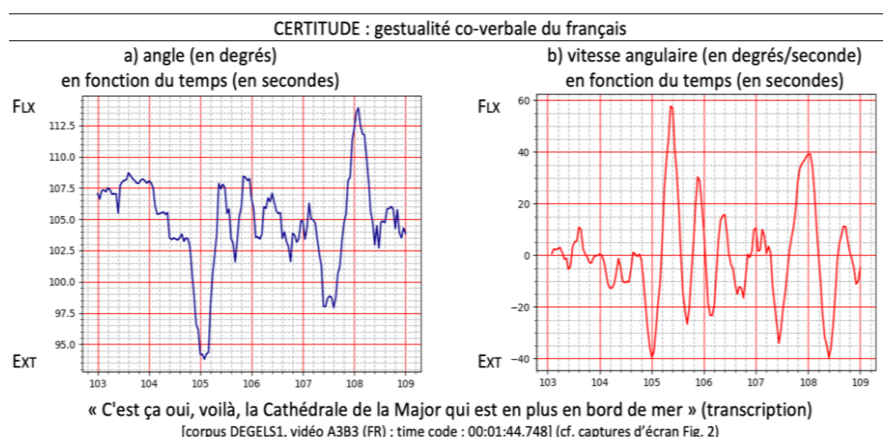

*Graphique 1 – Courbes montrant (a) les variations du DdL FlxExt du cou et (b) la vitesse de ces mouvements*

---

[15] Les étudiants en École d'Ingénieur Khodor Bou Orm (2022 ; M1 Polytech Sorbonne), Rabii Fath (2023 ; M2 Université de Rouen-Normandie) et Christine Chen (2024 ; M2 Polytech Sorbonne) qui, au cours de leur stage, ont programmé le logiciel permettant de visualiser les courbes, et Ludovic Saint-Bauzel, roboticien du laboratoire ISIR (Sorbonne Université) et coordinateur du projet LexiKHuM, qui a assuré le suivi technique de cette étape de travail.

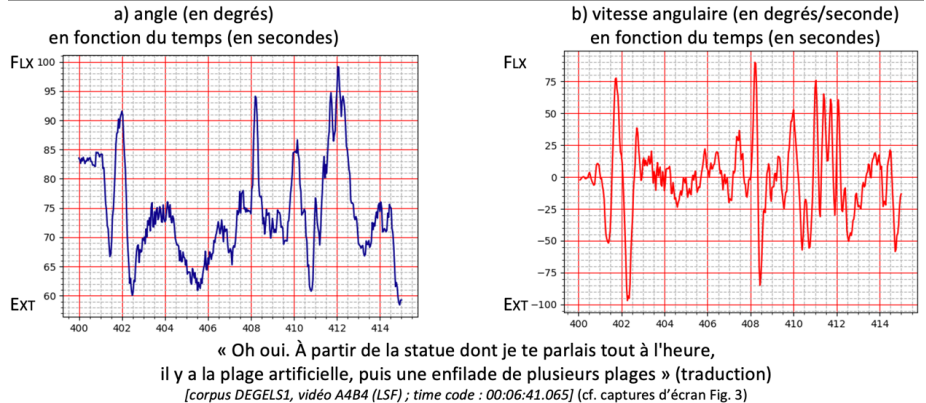

*Graphique 2 – Courbes montrant (a) les variations du DdL FlxExt du cou et (b) la vitesse de ces mouvements*

L'ordonnée de ces quatre courbes représente le temps de la séquence et est exprimé en secondes. Les deux courbes (a) représentent la position du cou sur l'axe du DdL de FlxExt qui est mesurée en degrés. Par exemple, le Graphique 1a montre que le sujet enregistré commence son énoncé par une tenue de la position du cou à un angle considéré comme la position neutre (soit la tête droite sans marque de FlxExt), puis, à la seconde 105, réalise un large mouvement d'Ext du cou, quasiment jusqu'à la butée articulaire, de la seconde 105 à 107,5, quatre mouvements FlxExt de fable amplitude, soit un acquiescement composé de quatre faibles hochements, et enfin une large Flx, quasiment jusqu'à la butée articulaire. Les deux courbes (b) montrent la vitesse des déplacements du cou sur l'axe FlxExt mesurée en degrés/seconde. Ainsi, le Graphique 2b montre que le sujet réalise un mouvement à une vitesse proche de 0°/s de la seconde 403 à 405, et donc que la position du cou est tenue (car sans vitesse de mouvement), puis une succession de hochements rapides (jusqu'à 60°/s) du cou.

Le recours aux données issues d'AlphaPose est très prometteur, mais il s'agit néanmoins d'une technologie en cours de développement. Il est important de noter qu'actuellement les mesures recueillies sont relatives à chaque vidéo et à chaque sujet enregistré. En effet, une étude préliminaire menée par notre équipe en décembre 2022[16], et non publiée, montre que les mesures réalisées avec AlphaPose dépendent (entre autres) de la distance du caméscope avec le sujet filmé, de son angle de prise de vue, de la morphologie du sujet et du *cadre*

---

[16] Avec la collaboration de Michael Nauge, ingénieur de recherche du laboratoire FoReLLIS de Poitiers.

*de détection*[17] sélectionné dans le logiciel de visualisation des données issues d'AlphaPose. La mise en place d'un protocole d'enregistrement uniforme, tel que celui réalisé par les concepteurs du corpus DEGELS1, permet de minimiser l'impact des différences de distances entre le camescope et le sujet, et à l'angle de vue ; l'extraction simultanée des données concernant chaque locuteur et chaque vidéo évite les problèmes potentiels associés au cadre de détection.

Les correctifs ci-dessus ne permettent toutefois pas, à eux tout seuls, d'obtenir des données numériques comparables : le Tableau 3 montre, par exemple, que la valeur d'EXT maximale pour le locuteur de français semble très proche de la valeur de FLX maximale du locuteur de LSF, ce qui semble contrintuitif.

| valeur numérique associée par AlphaPose pour : | locuteur de français (Graphique 1 et Figure 2) | locuteur de LSF (Graphique 2 et Figure 3) |
|---|---|---|
| valeur de la FLX maximale présente dans l'extrait analysé | 114° | 99,5° |
| valeur de l'EXT maximale présente dans l'extrait analysé | 94° | 60° |

*Tableau 3 – Valeurs numériques associées par AlphaPose aux valeurs maximales de FLX et EXT présentes dans les courbes des locuteurs de français (Figure 2) et de LSF (Figure 3)*

Puisque le protocole de DEGELS1 est uniforme, les divergences restantes entre les données sont en très grande partie dues à des différences morphologiques entre les locuteurs. Aborder ce problème, qui ne peut être résolu par une approche purement technique, impose de trouver une façon d'exprimer de manière homogène les données de différents locuteurs ; ce qui est rendu possible par la complémentarité entre le protocole d'annotation manuelle et celui d'annotation semi-automatique. En effet, il a été possible d'identifier manuellement différentes positions du sujet enregistré (notamment ses positions neutres et ses butées articulaires). Les courbes des séquences contenant des butées ou des positions de repos ont été annotées manuellement à l'aide de Typannot, qui fournit un codage de l'amplitude des différents DDL. Il a été ensuite possible d'associer (Tableau 4) une gamme de données angulaires aux crans de Typannot indiquant les butées (⌂ et ●), la position de repos (⌑) mais aussi les valeurs intermédiaires (grand ⌐ ⌐, moyen ⌐⌐, ou petit ⌐ ⌐).

---

[17] Dans le logiciel qui permet la visualisation des données issues de AlphaPose, avant toute opération de mesure, il faut tracer à main levée un cadre de détection autour du locuteur ; dans la première version du logiciel ce cadre ne pouvait être sauvegardé et les données devaient obligatoirement être extraites au même temps pour être comparable ; dans la dernière version, ce problème technique a été résolu et il est désormais possible de maintenir le même cadre (il faut toutefois penser à prendre note du cadre utilisé, car le logiciel ne le garde pas en mémoire, il faut réinitialiser les données à chaque fois). Toutefois, si deux vidéos n'ont pas été filmées en suivant de manière stricte le même protocole (notamment relativement à la distance du camescope ou à la qualité de l'image) la sauvegarde du cadre s'avère inutile.

|  | locuteur de français (Graphique 1 et Figure 2) | | locuteur de LSF (Graphique 2 et Figure 3) | |
|---|---|---|---|---|
|  | AlphaPose | Typannot | AlphaPose | Typannot |
| butée en FLX | 140° | ⊓⌂ | 137° | ⊓⌂ |
| valeur de la FLX maximale présente dans l'extrait analysé | 114° | ⊓⌐ | 99,5° | ⊓⌐ |
| position de repos | 104° | ⊓⌑ | 97° | ⊓⌑ |
| valeur de l'EXT maximale présente dans l'extrait analysé | 94° | ⊓◄ | 60° | ⊓◄ |
| butée en EXT | 88° | ⊓● | 53° | ⊓● |

*Tableau 4 – Correspondance entre les mesures d'AlphaPose (cf. Graphique 1 et Graphique 2) et le codage Typannot, pour différentes positions de la tête des locuteurs de français (Figure 2) et de LSF (Figure 3)*

En comparaison avec le codage réalisé à l'œil nu dans la première phase de notre méthodologie, l'utilisation des données issues de AlphaPose permet d'obtenir des annotations plus objectives et plus fines des données, rendant plus fiables les analyses. La conversion en Typannot de ces données, qu'elles soient le fruit d'une analyse à l'œil nu ou par MoCap/PosEst, assure un crantage homogène de toutes les courbes, permettant ainsi la comparaison entre les courbes de différents locuteurs mais aussi, si besoin, la comparaison de courbes ayant été filmées dans des conditions moins contrôlées que celles de production de DEGELS1.

Dans un avenir proche, le développement d'un algorithme convertissant directement les données issues de MoCap/PosEst en Typannot permettra d'accélérer le processus de transcription des données et facilitera l'extension des analyses menées à un nombre plus conséquent d'échantillons et/ou à un corpus plus vaste.

## 3. Analyses exploratoires

L'application du protocole de saisie et de traitement des données présenté ci-dessus nous a permis d'étudier les caractéristiques articulatoires des gestes épistémiques du cou chez des locuteurs de LSF et de français. Les analyses ont été conduites à partir des transcriptions en Typannot et des annotations relatives à la structure temporelle et rythmique associées aux changements de DDL.

Ci-dessous sont présentées les observations menées à partir de notre échantillon de 40 séquences épistémiques : 20 séquences de certitudes et 20 d'incertitude, subdivisées entre LSF et GCV de façon homogène. Les résultats obtenus ne sont que préliminaires et ne sont pas toujours significatifs[18], ils sont donc présentés à titre d'exemple afin de montrer le potentiel de la méthodologie mise en place.

La revue de littérature menée (voir § 1.2 et Tableau 1) indique que les hochements de la tête jouent un rôle pour marquer la certitude (par des hochements répétés, clairs et rapides, souvent appelés *nodding*) et l'incertitude (par des hochements lents). Ces hochements de la tête sont dus à des mouvements de FLX et/ou d'EXT du cou, avec une amplitude, une vitesse et un nombre de répétitions qui peut varier.

### 3.1. *FLXEXT du cou en certitude*

Sur les 20 séquences de certitude, le *nodding* a bien été observé tant en GCV (10/10) qu'en LSF (9/10) (Graphique 3). De ces FLXEXT du cou, 4/20 (1/10 en GCV et 3/10 en LSF) ont été réalisés avec un fort contraste d'amplitude et 13/20 (5/10 en GCV et 8/10 en LSF) ont traversé la position neutre (Graphique 4).

---

[18] Pour évaluer la qualité des résultats numériques, on peut utiliser le tableau ci-dessous, qui montre les espérances mathématiques d'occurrences cumulatives (positives ou négatives) après *n* répétitions (*p*=1-*q* avec *p*=*q*=0,5)

| répétions | 0 | 1 | 2 | 3 | 4 | 5 | 6 | 7 | 8 | 9 | 10 | 11 | 12 | 13 | 14 | 15 | 16 | 17 | 18 | 19 | 20 |
|---|---|---|---|---|---|---|---|---|---|---|---|---|---|---|---|---|---|---|---|---|---|
| 10 | 0.001 | 0.011 | 0.055 | 0.172 | 0.377 | 0.623 | 0.828 | 0.945 | 0.989 | 0.999 | 1.000 | | | | | | | | | | |
| 20 | 0.000 | 0.000 | 0.000 | 0.001 | 0.006 | 0.021 | 0.058 | 0.132 | 0.252 | 0.412 | 0.588 | 0.748 | 0.868 | 0.942 | 0.979 | 0.994 | 0.999 | 1.000 | 1.000 | 1.000 | 1.000 |

occurrences cumulatives

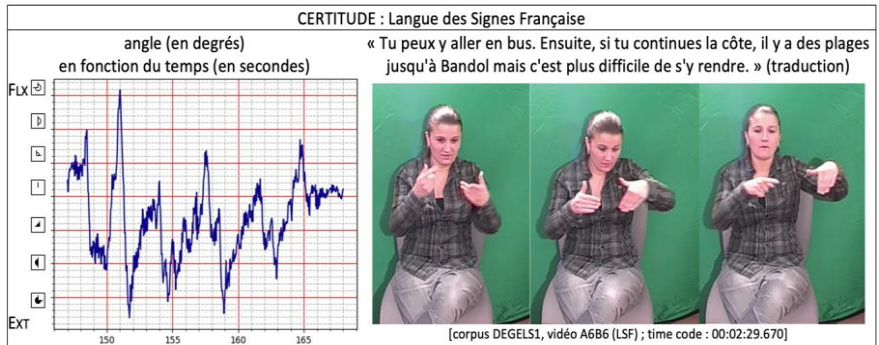

*Graphique 3 – Courbe de la position du cou montrant (cf. pics) une succession de mouvements de forte amplitude sur le DDL FLXEXT*

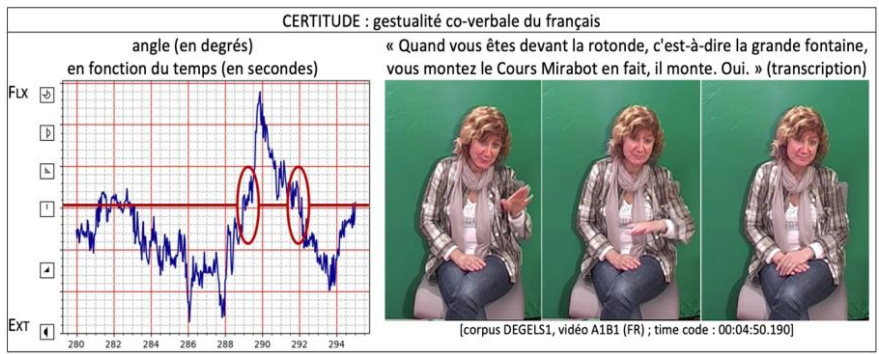

*Graphique 4 – Courbe de la position du cou montrant (cf. encadré) un mouvement sur le DDL FLXEXT traversant deux fois la position neutre*

Les séquences d'incertitude sélectionnées pour notre analyse ne présentent que très marginalement des hochements de tête (5/20), et ces hochements ne sont jamais répétés. Ils peuvent d'ailleurs souvent être interprétés comme des changements de position du cou, plus que comme des hochements de la tête.

Les observations de notre échantillon confirment que le hochement répété de la tête est un marqueur de certitude, mais – par rapport aux données issues de la littérature – une précision ultérieure est nécessaire : alors que le hochement avec un passage par la position neutre pourrait être considéré comme un marqueur de certitude, la seule forte amplitude du mouvement de FLXEXT du cou ne semble pas être un marqueur récurrent de la certitude.

### 3.2. *FLXEXT du cou en incertitude*

Dans l'échantillon étudié, 15/20 séquences d'incertitude (6 en GCV et 7 en LSF) montrent la présence d'une tenue – supérieure à deux secondes – sur l'axe

FLXEXT du cou (par exemple : Graphique 5). Parmi ces 15 tenues, 10 n'ont pas été réalisées en position neutre (le cou avait donc une FLX ou une EXT marquée).

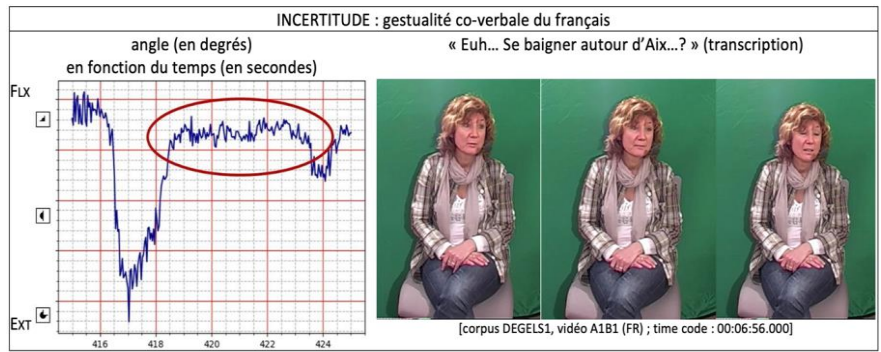

*Graphique 5 – Courbe de la position du montrant (cf. encadré) une tenue (>2 sec) de la position sur le DDL FLXEXT*

L'observation des séquences de certitude, à l'inverse, montre la quasi-absence de tenue (2/20, réalisées, toutes deux, proches de la position neutre). Notons aussi que deux séquences en GCV n'ont pas été comptabilisées car elles ne présentaient pas une tenue nette, mais plutôt des micro-oscillations. Ces dernières auraient pu être considérées comme des micro-oscillations physiologiques naturelles en tenue de position (et donc être ajoutées aux 15/20 tenues identifiées).

Dans la littérature, l'incertitude n'est pas associée à une position spécifique de la tête. Toutefois, les données issues de l'analyse de notre échantillon montrent que l'incertitude peut être véhiculée par une tenue (pouvant durer plusieurs secondes) de la tête en position de FLX ou d'EXT (c'est-à-dire en dehors de la position neutre).

### 3.3. *Amplitude et vitesse de déplacement du cou en certitude et incertitude*

Deux mouvements du même DDL d'un même articulateur parcourant la même distance mais à vitesse différente présentent des caractéristiques cinématiques différentes : visuellement, le mouvement plus rapide semble plus tendu, et le mouvement plus lent semble plus relâché. Pour pouvoir interpréter la tension d'un mouvement, il faut observer la courbe de vitesse du mouvement, ainsi que la courbe représentant les variations de position du segment, qui montre sa distance parcourue selon un DDL donné. Les mesures de vitesse (exprimées en degrés/seconde, °/s) peuvent être obtenues grâce à AlphaPose et sont relatives à chaque sujet.

Les premières observations montrent que 15/20 des séquences (6/10 en GCV et 9/10 en LSF) de notre échantillon présentent une vitesse élevée des mouvements de FlxExt du cou en certitude (plus de 40°/s, comme dans le passage encadré du Graphique 6) et que 15/20 (10/10 en GCV et 5/10 en LSF) des séquences d'incertitude ont une moindre vitesse (moins de 20°/s comme dans le passage encadré du Graphique 7). À l'inverse, dans les séquences d'incertitude les échantillons présentent plus rarement des mouvements rapides (3/20) et, en certitude, ceux présentant des mouvements lents (2/20).

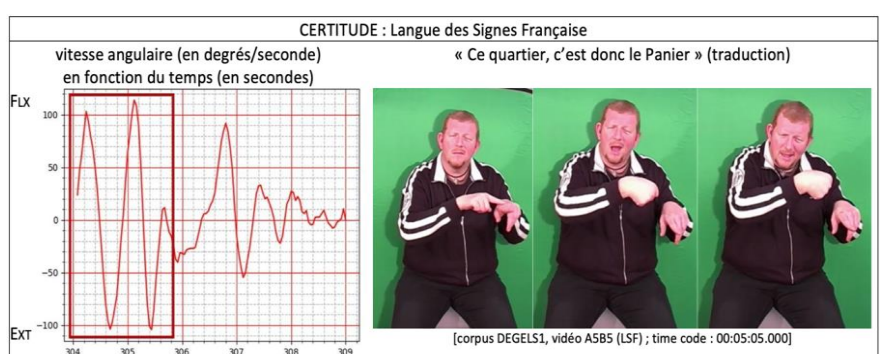

*Graphique 6 – Courbe de la vitesse du mouvement du cou sur le DdL FlxExt montrant (cf. encadré) une vitesse élevée (>40°/s)*

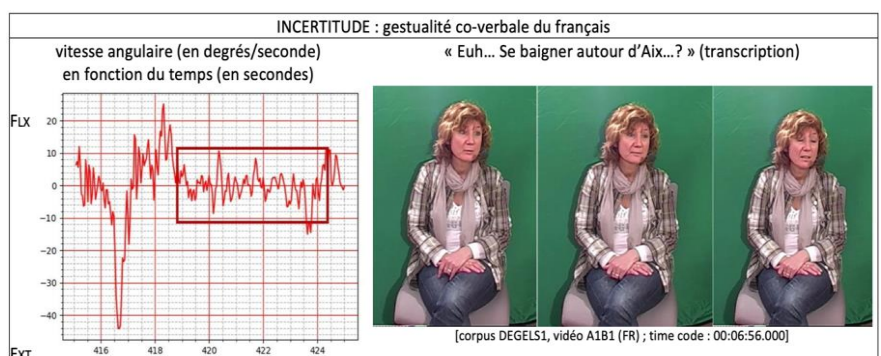

*Graphique 7 – Courbe de la vitesse du mouvement du cou sur le DdL FlxExt montrant (cf. encadré) une faible vitesse (<20°/s)*

Au vu des mesures d'amplitude et de vitesse des mouvements épistémiques relevées dans cette étude, il serait possible de considérer que les mouvements de certitude sont réalisés et perçus comme plus tendus et les mouvements d'incertitude plus relâchés. Néanmoins, une vérification de l'amplitude absolue de chaque mouvement est nécessaire pour pouvoir l'affirmer (ce qu'AlphaPose ne permet pas encore).

4.     Conclusion générale et perspectives

Dans cet article, une vaste place est réservée à la présentation de la méthodologie élaborée pour décrire et analyser les mouvements corporels en GcV et en LS. L'utilisation combinée d'une annotation manuelle et d'une annotation semi-automatique issues de AlphaPose, toutes deux associées à une transcription en Typannot, a permis d'obtenir une description articulatoire affinée, mesurée et plus objective des données.

L'élaboration de méthodes pour valider les données issues de MoCap/PosEst contribue à poser les bases pour une annotation semi-automatique des données concernant la forme des gestes et de la LS. Cette avancée permettra de créer de vastes corpus transcrits de manière plus rapide, facilitant ainsi l'étude quantitative des relations liant la forme et le sens en GcV et LS.

L'application de cette méthodologie d'extraction et analyse des données au corpus DELEGS1 a permis d'identifier les caractéristiques articulatoires, communes aux GcV et aux LS, des mouvements présents dans la gestualité épistémique. Dans le cadre plus vaste du projet LexiKHuM, ce travail permet de repérer les traits articulatoires qui pourraient être implémentés dans un dispositif haptique véhiculant des messages épistémiques. Le SI auquel sera lié ce dispositif pourra ainsi informer l'opérateur humain sur le statut des informations qu'il fournit, diminuant de faite son opacité et améliorant l'interaction humain-machine.

Le protocole développé a permis d'analyser de manière homogène des données issues de la GcV du français et de la LSF à la recherche des caractéristiques communes à ces deux modalités. Bien que les résultats présentés ici soient préliminaires, ils montrent que (1) des hochements de tête passant par la position neutre et une vitesse de mouvement élevée sont des marqueurs de certitude et (2) les tenues de position de la tête hors de la position neutre et une faible vitesse de mouvement indiquent l'incertitude. Par rapport aux travaux citées en littérature, la recherche conduite propose une analyse plus détaillée de la gestualité épistémique : ainsi, la certitude n'est pas uniquement véhiculée par des « hochements clairs et rapides de la tête » mais par des mouvements *répétés* de *FLX et EXT du cou* qui sont caractérisés par une *vitesse élevée* mais aussi par le fait que *la variation de DDL traverse la position de repos*.

À court terme, nous envisageons de continuer nos recherches le long de trois axes : augmenter le nombre d'échantillons relatifs à la FlxExt du cou afin de valider (ou pas) statistiquement nos résultats préliminaires ; vérifier que les marqueurs identifiés ne soient présents qu'en situation d'épistémicité et donc qu'il s'agisse bien des marqueurs de certitude ou incertitude ; continuer l'exploration des gestes épistémiques en l'étendant à d'autres DdL (l'AbdAdd et la RinRex) et à d'autres articulateurs (notamment les épaules et le buste), et en y associant aussi l'analyse du regard. Il est aussi pris en considération d'étendre la recherche à un corpus plus vaste et varié que le seul DEGELS1. Sur le long terme, notre objectif est de ne plus traiter chaque articulateur de manière individuelle, mais voir si des corrélations/combinaisons entre DdL d'un même ou de différents segments sont pertinentes. Cela permettrait de mettre en évidences des caractéristiques du mouvement qui véhiculerait la certitude ou l'incertitude indépendamment du segment mobilisé dans le geste épistémique.

Lorsque les caractéristiques articulatoires des gestes épistémiques seront identifiées, les membres du projet comptent explorer d'autres unités de sens qui pourraient être véhiculées par le SI afin d'améliorer la transmission de ses intentions à son opérateur humain : l'expression de la dangerosité et de l'urgence. Pour cela, il est envisagé d'utiliser le même protocole que celui mis en place pour étudier l'épistémicité, mais en l'appliquant à un autre corpus. En étendant le vocabulaire kinesthésique développé dans le cadre du projet, il sera possible de combiner différents types de messages, pour permettre au pilote d'hélicoptère de connaître le degré de certitude de son système intelligent par rapport à l'information donnée, mais aussi d'évaluer la rapidité avec laquelle il doit réagir à cette information (urgence) et la gravité des conséquences d'un manque de réaction ou d'une réaction erronée de sa part (dangerosité).

**Crédits et remerciements**

**Bibliographie**

Bianchini C. S., Chevrefils L., Danet C., Doan P., Rébulard M., Contesse A. & Boutet D., 2018, Coding movement in Sign Languages: the Typannot approach, *Proceedings of the 5th International Conference on Movement and Computing* [*MoCo'18*], Genova, ACM, sect. 1(9), p. 1-8.

Bianchini C. S., 2024, (*D*)*écrire les langues des Signes : approche grapholinguistique aux Langues des Signes*, Brest, Fluxus Éditions.

Borràs-Comes J., Roseano P., Vanrell M., Chen A. & Prieto P., 2011, Perceiving uncertainty: facial gestures, intonation, and lexical choice, *Proceedings of the 2nd Conference on Gesture and Speech in Interaction* [*GESPIN2011*], Bielefeld.

Borràs-Comes J., Kiagia E. & Prieto P., 2019, Epistemic intonation and epistemic gesture are mutually co-expressive: empirical results from two intonation-gesture matching tasks, *Journal of Pragmatics* 150, p. 39–52.

Boutet D., 2008, Une morphologie de la gestualité : structuration articulaire, *Cahiers de Linguistique Analogique* 5, p. 81-115.

Boutet D., 2018, *Pour une approche kinésiologique de la gestualité*, Habilitation à diriger des recherches [HDR], Université de Rouen Normandie.

Boutet D., Blondel M., Beaupoil-Hourdel P. & Morgenstern A., 2021, A multimodal and kinesiological approach to the development of negation in signing and non signing children, *Languages and Modalities* 1(1), p. 31-47.

Boyes-Braem P., 1999, Rhythmic temporal patterns in the signing of deaf early and late learners of Swiss German Sign Language, *Language and Speech* 42, p. 177–208.

Braffort A. & Boutora L., 2012, DEGELS1: a comparable corpus of French Sign Language and co-speech gestures, *Proceedings of the 8th International Conference on Language Resources and Evaluation* [*LREC'12*], Paris, ELRA, p. 2426-2429.

Bross F., 2012, German modal particles and the common ground. *Helikon: a Multidisciplinary Online Journal* 2, p. 182 –209.

Catteau F. 2020, *Traduire la poésie en langue des signes : l'empreinte prosodique lors du changement de modalité*, Thèse de doctorat, Université de Paris 8.

Chevrefils L., Danet C., Doan P., Thomas C., Rébulard M., Contesse A., Dauphin J.-F. & Bianchini C. S., 2021, The body between meaning and form: kinesiological analysis and typographical representation of movement in Sign Languages, *Languages and Modalities* 1(1), p. 49-63.

Chevrefils L., 2022, *Formalisation et modélisation du mouvement en Langue des Signes Française : pour une approche kinésio-linguistique des productions gestuelles*, Thèse de doctorat, Université de Rouen Normandie.

Christoffersen K. & Wood D., 2002, How to make automated systems team players, *Advances in Human Performance and Cognitive Engineering Research* 2, Bingley, Emerald, p. 1-12.

Crasborn O. & Sloetjes H., 2008, Enhanced ELAN functionality for sign language corpora, *Proceedings of the 3rd Workshop of the 6th International Conference on Language Resources and Evaluation [LREC'08]*, Paris, ELRA, p. 39-43.

de Haan F., 2005, Encoding speaker perspective: evidentials, *Linguistic diversity and language theories*, Amsterdam: John Benjamins, p. 379-417.

Debras C., 2017, The shrug: forms and meanings of a compound enactment, *Gesture* 16(1), p. 1-34.

Dekker S. W. A. & Woods D. D., 2002, MABA-MABA or abracadabra? Progress on human-automation co-ordination, *Cognition, Technology & Work* 4, p. 240–244.

Fang H.-S., Li J., Tang H., Xu C., Zhu H., Xiu Y., Li Y.-L. & Lu C., 2023, AlphaPose: whole-body regional multi-person pose estimation and tracking in real-time. *IEEE Transactions on Pattern Analysis and Machine Intelligence* 45(6), p. 7157-7173.

Ferré G., 2012, Functions of three open-palm hand gestures, *Multimodal Communication* 1, p. 5-20.

Gianfreda G., Volterra V. & Zuczkowski A., 2014, L'espressione dell'incertezza nella Lingua dei Segni Italiana (LIS), *Ricerche di Pedagogia e Didattica - Journal of Theories and Research in Education* 9(1), p. 199-234.

Grynszpan O., Sahaï A., Hamidi N., Pacherie E., Berberian B., Roche L. & Saint-Bauzel L., 2019, The sense of agency in human-human *vs* human-robot joint action, *Consciousness and Cognition* 75, #102820, p. 1-14.

Guillain A. & Pry R., 2012, D'un miroir l'autre: fonction posturale et neurones miroirs, *Bulletin de Psychologie* 518(2), p. 115-127.

Herrmann A. *Modal and focus particles in Sign Languages: a cross-linguistics study*, Berlin, De Gruyter Mouton.

Jantunen T., Mesch J., Puupponen A. & Laaksonen J., 2016, On the rhythm of head movements in Finnish and Swedish Sign Language sentences, *Proceedings of the 8th International Conference on Speech Prosody*, Boston, ISCA, p. 850–853.

Karabüklü S., Bross F., Wilbur R. & Hole D., 2018, Modal signs and scope relations in TİD, *Proceedings of Formal and Experimental Advances in Sign Language Theory [FEAST]* 2, p. 82-92.

Kronning H. 2012, Le conditionnel épistémique : propriétés et fonctions discursives, *Langue Française* 173, p. 83-97.

Martel K., Dodane C., Blondel M., Catteau F. & Bellifemine C., 2022, Processus de focalisation multimodale lors des dîners familiaux français : une comparaison entre familles parlantes et familles signantes, *Faits de Langues*, dans ce même volume.

Pezzullo G., Roche L. & Saint-Bauzel L., 2021, Haptic communication optimises joint decisions and affords implicit confidence sharing, *Scientific Reports* 11, #1051.

Puupponen A., Wainio T., Burger B. & Jantunen T., 2015, Head movements in Finnish Sign Language on the basis of motion capture data: a study of the form and function of nods, nodding, head thrusts, and head pulls, *Sign Language & Linguistics* 18(1), p. 41-89.

Roseano P., González M., Borràs-Comes J. & Prieto P., 2016, Communicating epistemic stance: how speech and gesture patterns reflect epistemicity and evidentiality, *Discourse Processes* 53, p. 135-174.

Taleb N. N., 2012, *Antifragile: how to live in a world we don't understand*, London, Allen Lane.

Thomas C., en course, *Étude des paramètres non-manuels en LSF au sein d'énoncés interrogatifs : entre transcriptions manuelles et capture de mouvement*, Thèse de doctorat, Université de Rouen Normandie.

Wilbur R. B., 1999, Stress in ASL: empirical evidence and linguistic issues, *Language and Speech* 42(2), p. 229–250.

Wilbur R. B. & Malaia E., 2018, A new technique for analyzing narrative prosodic effects in sign languages using motion capture technology, *Linguistic foundations of narration in spoken and sign languages*, Amsterdam, John Benjamins, p. 15–40.

**Liste des abréviations**

| | |
|---|---|
| ABD | Abduction |
| ADD | Adduction |
| ANR | Agence Nationale de la Recherche |
| DDL | Dégrées de liberté |
| DEGELS1 | DEfi Geste Langue des Signes |
| DGS | Langue des signes allemande (Deutsche Gebärdensprache) |
| DTIS | Département de Traitement de l'Information et Systèmes |
| EXT | Extension |
| FLX | Flexion |
| FoReLLIS | Formes et Représentations en Linguistique, Littérature et dans les arts de l'Image et de la Scène (UR15076) |
| FR | Français oral |
| GCV | Gestualité co-verbale |
| ISIR | Institut des Systèmes Intelligents et de Robotique (UMR 7222) |
| LexiKHuM | Lexique d'interaction Kinésiologique Humain-Machine |
| LIS | Langue des signes italienne (Lingua dei segni italiana) |
| LISN | Laboratoire Interdisciplinaire Sciences du Numérique (UMR 9015) |
| LS | Langue des signes |
| LSF | Langue des signes française |
| LV | Langue vocale |
| M1 / M2 | Master 1, Master 2 |
| MOCAP | Motion capture |
| ONERA | Office national d'études et de recherches aérospatiales |
| POSEST | Estimation de pose |
| REX | Rotation externe |
| RIN | Rotation interne |
| SI | Système Intelligent |
| TİD | Langue des signes turque (Türk işaret dili) |